\documentstyle[12pt]{article}
\begin{document}
\title{
           Is 2d Turbulence a Conformal Turbulence?}
\author 
{Gregory Falkovich\cite{NSK} and Amihay Hanany}
{\em
Physics Department, Weizmann Institute of Science,
Rehovot 76100, Israel
}
\date{\today}
\maketitle

\begin{abstract}
A critical analysis of the conformal approach to the theory of 2d
turbulence is delivered. It is shown, in particular, that conformal
minimal models cannot give a general turbulent solution,
which should provide for constant fluxes of all vorticity integrals of
motion.
\end{abstract}
\overfullrule=0pt
The problem of small-scale spectrum of two-dimensional turbulence is a
peculiar problem among the variety of turbulent systems. The point is
that dimensional considerations do not give a steady spectrum that
corresponds to the vorticity cascade. The spectrum
obtained from dimensional analysis is $E(k)\propto k\sp{-3}$
\lbrack1\rbrack\ which yields a logarithmic infrared divergence after
substitution into the equations for the correlation functions.
Kraichnan's attempt to save the spectrum from nonlocality
by introducing the slow factor $\ln\sp{-1/3}k$ attains convergence only
in the first order of perturbation theory \lbrack2\rbrack\
while the next orders reveal
divergences with higher powers of the logarithm: $\ln\sp2$ etc. The fact
that the powers of the logarithm increase with the order of perturbation
theory suggests that a substantial renormalization of the index occurs.
No successful attempts to work out the divergences or to show that
they are cancelled are known to us. The existence of
alternative predictions for the steady spectrum $E(k)\propto k\sp{-4}$
by Saffman \lbrack3\rbrack\ and $E(k)\propto k\sp{-11/3}$ by Moffatt
\lbrack4\rbrack, show that this is still an open problem.

A fairly new approach to the problem has recently been
introduced by Polyakov \lbrack5\rbrack. He suggested to borrow a set of
correlation functions from conformal field theory to satisfy
a chain of equations following from Euler's equation.
Conformal invariance is assumed for
the set of the simultaneous correlation functions
(not for the Euler equation itself). This is based on
the idea that nonlinear interaction
renormalizes all the properties in such a way that at criticality
(for an infinite Reynolds number, in our case) system possesses the
highest possible symmetry (absent in the initial equation).
It is convenient to start from the Navier-Stokes equation written
for the vorticity field $\omega({\bf x},t)$
\begin{equation}\dot\omega+e_{\alpha\beta}\partial_\alpha\psi
\partial_\beta\Delta\psi=\nu\Delta\omega\ ,\label{Euler}\end{equation}
$\psi$ is a stream function giving the velocity field as
$v_\alpha=e_{\alpha\beta}\partial_\beta\psi$. Here and below we use
the notation $\partial/\partial x_\alpha=\partial_\alpha$.

Our aim is to find a stationary set of equal time correlation functions
$I_n({\bf x}_1,\ldots,{\bf x}_n)=\langle
\omega({\bf x}_1,t)\ldots \omega({\bf x}_n,t)\rangle$.
The brackets denote an average with some time independent probability
distribution:
$$\sum_{p=1}\sp n
\left\langle\omega({\bf x}_1,t)\ldots\dot\omega(
{\bf x}_p,t)\ldots\omega({\bf x}_n,t)\right
\rangle=0\ .$$
Such a stationary set is expected to exist in the inertial interval of
scales, i.e., for distances that are much less than the scale $L$ of an
external pump
and much larger than the viscous scale $a$. It is possible, then, to
neglect viscosity in (1) using instead a careful point splitting
procedure \lbrack5\rbrack\ for the nonlinear term
$$
e_{\alpha\beta}\partial_\alpha\psi({\bf x})\partial_\beta\Delta\psi({\bf
x})=\!\!\lim_{a\rightarrow0}e_{\alpha\beta}\partial_\alpha\psi({\bf x}+{
\bf a})\partial_\beta\Delta\psi({\bf x}-{\bf a})$$
where ``lim'' implies angle averaging. To calculate different-point pair
correlators like $\psi({\bf x}+{\bf a})\psi({\bf x}-{\bf a})$,
the fusion rule of the type
\begin{equation}
\lbrack\psi\rbrack\,\lbrack\psi\rbrack=\lbrack\phi\rbrack+\ldots
\label{ope}\end{equation}
should be used.
Here we follow the notations of Ref.5 so that $\lbrack\psi
\rbrack$ means the conformal class of $\psi$, i.e. itself together
with the operators $L_{-n_1}\ldots L_{-n_k}\psi$, $L_{-n}$ being
Virasoro generators \lbrack6\rbrack. Both $\psi$ and $\phi$ are presumed
to be taken from a set (primary fields) of some conformal field theory.

Polyakov suggested to use the so called minimal models
\lbrack6\rbrack\  which contain a finite number of primary fields.
The main subject of this paper is to show that a minimal model
cannot serve as a general turbulent solution.

The primary field $\phi$ provides
the main contribution in the operator product expansion (OPE)
(\ref{ope}) in the small-scale region, i.e. it has the smallest conformal
dimension. The important thing is that the scaling indices (dimensions)
of the fields are not additive so generally
$\Delta_\phi\not=2\Delta_\psi$ and a dimension defect appears.
The energy density in the wave number space is expressed via $\vert\psi
\vert\sp2$ and is
\begin{equation}E(k)\propto k\sp{4\Delta_\psi+1}\ ,
\label{ener}\end{equation}

To choose an appropriate solution from the wealth of conformal solutions,
one should impose some additional conditions that follow from the
symmetries or conservation laws specific for the problem in question.
According to Fjortoft's theorem (see e.g. \lbrack7\rbrack),
the vorticity is the relevant quantity in the problem of small-scale
turbulence (while the energy flux determines large-scale turbulence).
Following Kraichnan \lbrack1\rbrack\ who developed a simple and efficient
(though uncontrollable, of course) closure in terms of double correlation
functions, the enstrophy
$$H_2=\int\omega\sp2({\bf x})d\sp2x\ ,$$
which is a motion integral of Euler's equation is usually taken into
account.
A steady turbulence spectrum in the small-scale region should provide for
a constant enstrophy flux over the scales which yields \lbrack5\rbrack\
\begin{equation}
\left\langle\dot\omega(x+r)\omega(x)\right\rangle
\propto r\sp0={\rm const}\ .\label{flux}\end{equation}
Puting $\omega=\Delta\psi$ and
$\dot\omega\propto \bigl(
L_{-2}\bar L_{-1}\sp2-\bar L_{-2}L_{-1}\sp2\bigr)\phi$,
Polyakov obtained \lbrack5\rbrack\
\begin{equation}
(\Delta_\phi+2)+(\Delta_\psi+1)=0\ .\label{fluxdim}\end{equation}
As one can see, the enstrophy flux is expressed through the triple
correlation function which can be expressed by the fusion
rule (\ref{ope}) in terms of the double correlation function.

Equation (\ref{fluxdim})
can be obtained also by requiring the rate of the enstrophy
dissipation to remain constant while the viscosity $\nu$ goes to zero:
\begin{equation}{dH_2\over dt}=\nu\int\sp{1/a}k\sp2H_kdk
\propto\nu a\sp{-6-4\Delta_\psi}
\propto\nu\sp{3+\Delta_\phi+\Delta_\psi\over\Delta_\phi-
\Delta_\psi}\,.\label{dissip}
\end{equation}
The last estimate was given by using the expression for the
viscous scale $\nu\propto a\sp{2\Delta_\psi-2\Delta_\phi}$
that follows from the comparison of the nonlinear and the viscous terms
in the Navier-Stokes equation. The natural physical assumption that
the main dissipation stems from small scales is valid if
$\Delta_\psi>-3/2$
or $\Delta_\psi>\Delta_\phi$ as it can be seen from
(\ref{fluxdim},\ref{dissip}).

Note that
Kraichnan's dimensional approach would correspond to additive dimensions
$\Delta_\phi=2\Delta_\psi$ giving thus $\Delta_\psi=-1$ and $E(k)\propto
k\sp{-3}$. However, we have arguments that suggest that this in not a
conformal solution from the set of minimal models (see Appendix).

To ensure that the conformal set of correlators is a steady solution,
Polyakov imposed an extra condition requiring
that rhs of (\ref{ope}) (which determines the time derivatives of the
correlators) vanishes in the ultraviolet limit. Since
$\psi(z_1)\psi(z_2)=(z_1-z_2)\sp{\Delta_\phi-2\Delta_\psi}
\phi(z_2)+\ldots$
as $z_1\rightarrow z_2$, the following inequality arises:
\begin{equation}\Delta_\phi>2\Delta_\psi\ ,\label{regul}\end{equation}
thus, using (\ref{fluxdim}), $\Delta_\psi<-1$.

Conditions (\ref{fluxdim}) and (\ref{regul})
by no means determine a single solution.
 The minimal model (2,21) presented by Polyakov \lbrack5\rbrack\ is
nothing but the first example from an infinite family. The curious reader
can find the first few hundred solutions in our preprint \cite{FH92}.
Polyakov's solution corresponds to the minimal number of primary
fields (in this case 10). One could, in particular, find the minimal
model (5,72) that gives $\Delta_\psi=\Delta_{(1,25)}=-7/6$ and
$E(k)\propto k\sp{-11/3}$ as in Moffatt's spectrum.

Usually, when speaking about a turbulent solution carrying a constant
flux, one should check that two conditions are satisfied: i) The solution
should be local in $k$-space which means that distant scales do not
interact substantially; this should be provided by the convergence of the
integral determining the flux in $k$-space;
ii) The constant that arise
in this (converging) integral should be nonzero and have the correct sign
to satisfy the boundary conditions in $k$-space, i.e.,
the pumping and damping. Any solution of (\ref{fluxdim}) and
(\ref{regul}) violates both of these conditions.

Inequality (\ref{regul}) means that
$\Delta_\psi<-1$ i.e. the spectrum (\ref{ener}) is steeper than
Kraichnan's one (which yields an infrared logarithmic divergence) so that
a power infrared divergence arises for any solution.
To show this, let us consider the Euler equation in the momentum
representation:
$$
{\partial \omega\sb k\over\partial t}
=\int\Gamma\sb {12}\omega\sb1\sp*\omega\sb2\sp*
\delta({\bf k}+{\bf k}\sb1+{\bf k}_2)\,d\sp 2 k\sb1 d\sp 2k\sb2\ .
$$
Here the vertex $\Gamma\sb{12}=\lbrack {\bf k}\sb1{\bf k}\sb2\rbrack
\bigl(k\sb1\sp{-2}-k\sb2\sp{-2}\bigr)$.
The equation for the enstrophy density
$h_2({\bf k},t)\delta({\bf k}+{\bf k}')=\langle\omega_k\omega_{k'}
\rangle$ by virtue of the conservation law can be written as a
continuity equation
$\dot h_2+{\rm div}\,{\bf P}\sb2=0$.
The flux is constant in a steady state:
\begin{equation}
P\sb2=2\pi\!\int\limits_0\sp kk\,dk\int \Gamma\sb {12}\langle\omega\sb k
\omega\sb1\omega\sb2\rangle
\delta({\bf k}+{\bf k}\sb1+{\bf k}_2)\,d\sp 2 k\sb1 d\sp 2k\sb2\,.
\label{fc2}
\end{equation}
This is the Fourier transform of (\ref{flux}).
Infrared convergence of (\ref{fc2}) is determined by the
asymptotics of the triple correlation function
at $k\sb1\ll k\simeq k\sb2$. Such an asymptotics is expressed
via the second-order correlation function \cite{LF92}:
\begin{equation}
\langle\omega\sb k\omega\sb1\omega\sb2\rangle=J_{k12}
h_2(k_1)\Phi(k)(k\sb1/k)\delta({\bf k}+{\bf k}\sb1+{\bf k}\sb2)\ .
\label{as3}
\end{equation}
Here $\Phi(k)$ is a power function which
value is irrelevant to the
problem of infrared convergence. Substituting (\ref{as3}) into
(\ref{fc2}), one gets the following integral
\begin{equation}\int\sb{1/L}kh_2(k)dk\ .\label{int3}\end{equation}
If $h_2(k)\propto k\sp{-y_2}$, then (\ref{int3}) converges for $y_2<2$.
It means that the energy spectrum $E(k)$ is local if it is less steeper
than Kraichnan's $k\sp{-3}$. For the conformal solutions,
$y_2=-2\Delta_\psi$
and the convergence condition gives $\Delta\sb\psi>-1$.
Equation (\ref{fc2}) gives the scaling exponent of the triple
correlation function $y_3=4$ which corresponds to (\ref{fluxdim}) only
in the case of convergence.

As far as the second condition (of nonzero flux) is concerned, the
three-point function $\langle\psi\psi\psi\rangle\sim\langle\phi\psi
\rangle$ is equal to zero since primary fields with different dimensions
($\Delta_\psi\not=\Delta_\phi$) are orthogonal for minimal models. The
cases where $\psi$ appears in the operator product expansion of $\lbrack
\psi\rbrack\lbrack\psi\rbrack$ (like those with $\Delta_\psi=\Delta_\phi=
-3/2$ which one can find in \cite{FH92}) break parity and therefore
should be excluded. One can consider models with $\Delta_\psi=\Delta_\phi
=-3/2$ but $\lbrack\psi\rbrack
\lbrack\psi\rbrack$ does not contain $\psi$ in the OPE, this solution
corresponds to the spectrum $E(k)\propto k\sp{-5}$ which seems to be
unphysical since the dissipation integral (\ref{dissip}) is not defined
by the ultraviolet region but stems from the whole $k$-space implying
thus the absence of the inertial interval in this case.

The conformal solutions in question are thus fluxless. This is quite
natural since they are invariant under time reflection while a nonzero
flux corresponds to an irreversible dissipation. The second
difficulty (zero flux) remedies to some extent the first difficulty
(nonlocality) since one should not require the convergence of the
integrals that are identically zero. Physically this corresponds to the
fact that the spectrum of a system in thermodynamic equilibrium should
not be local, unlike the cascade spectrum. A fluxless spectrum
corresponds to an equilibrium case.

But, since we are discussing a nonequilibrium situation, there
still remains the question, how the spectrum carries nonzero flux from
the pumping region to the viscous region of scales. Polyakov's suggestion
is that small (nonconformal) deviations from the power law due to
infrared cut-off (pumping scale) could provide nonzero values for the
enstrophy flux. (It is unclear if such a nonlocal yet cascade solution
exists until it is explicitly found by solving the matching problem).
The physical correlators are thus assumed to be close to the equilibrium
ones in the inertial interval of scales. This is similar to what happens
in two-dimensional optical turbulence described by the Nonlinear
Schr\"odinger Equation
$i\Psi_t+\Delta\Psi+T\vert\Psi\vert\sp2=0$.
For wave turbulence, the small-scale spectrum is $\epsilon(k)\propto
k\sp{\alpha-m-d}$, where $d$ is the space dimension and $\alpha$ and $m$
are the scaling indices of the Hamiltonian coefficients (i.e. the
frequency and the four-wave interaction coefficient respectively)
\cite{ZLF2}. For the NSE, $\alpha=d=2$ and $m=0$ so that the turbulent
spectrum is $\epsilon(k)=\,$const which coincides with
the equilibrium equipartition. This spectrum (which is an exact steady
solution) is fluxless too. Numerical simulations of the NSE show the
nonequilibrium spectrum to be close to $\epsilon\approx$ const
\cite{DNPZ}. The same coincidence of the equilibrium and
turbulent spectra takes place for common turbulence of Langmuir
and ion sound waves in plasma, the spectra carrying fluxes acquire
logarithmic factors in this case \cite{ZLF2}.

In our opinion, unlike the above cases of wave turbulence,
conformal approach, namely,
the assumption that the turbulent spectrum of 2d hydrodynamics should
be close to an equilibrium one, is not based on solid ground.
One would like to see the degeneracy that prescribes the coincidence
of the turbulence spectrum with an equilibrium one.

Polyakov suggested to distort the spectra by analytical (in
${\bf x}$-space) contributions. It corresponds to introducing
$\delta(k)$-terms in the momentum representation \cite{pol2}.
As a result, the flux (\ref{fc2}) is determined by $\delta$-terms
and it is thus automatically independent of $k$. To get the condition
(\ref{fluxdim}), Polyakov required that the flux is independent
of the infrared cut off which is the pump scale. This condition is
satisfied in the (purely theoretical) case of the excitation by
an external force independent of the velocity field. If one consider,
however, a more physical case of a large-scale instability
($\partial h_2/\partial t+{\rm div}\,P_2=\gamma(k)h_2$), then the
flux depends on the value of $h_2(k_0)$ at the pump region.
If one estimates this value by extending the power
solution from the inertial interval $h_2(k)=P\sp{a}k\sp{-s}$,
one has \cite{FS89}
$$P\propto k_0\sp{(1-s)/(1-a)}\ .$$
What is important here is the factor $1-s$. It
has clear physical meaning: if one considers the integral of the total
enstrophy $H=\int h_2(k)dk$, then $1-s$ determines whether
the enstrophy-containing region is in small or large scales.
The enstrophy flux increases if one shifts or extends the pump toward
the enstrophy-containing scale. Conformal solutions correspond to $s>1$.
There are thus three possibilities:
1) conformal solution cannot be matched with an instability
increment; 2) the value
$h(k_0)$ is much less than one naively estimated by extending the
conformal spectrum; 3) $h(k_0)$ is well matched with a conformal
part but into a pump integral you should substitute a much smaller
value. Maybe 3) means that actually only small
number of eddies grow due to the instability
since most of large eddies are quasisteady and do not
provide an instability leading to the direct cascade, while the
whole set of eddies constitutes the conformal solution.

And what is more important, the above conformal approach
does not account for the presence of an infinite set of motion integrals
\begin{equation}
H_n=\int\omega\sp{n}({\bf x})d\sp2x\ .\label{integ}\end{equation}
The conservation of $H_n$ follows directly from the fact that the
equation
$$\dot\omega+e_{\alpha\beta}\partial_\alpha{\delta{\cal H}\over\delta
\omega}\partial_\beta\omega=0$$
conserves the integral $\int F(\omega)\,dxdy$, where $F$ is an arbitrary
function and the Hamiltonian ${\cal H}$ is an arbitrary functional of
$\omega$ \lbrack not only ${\cal H}=\int\psi\omega\,dxdy$, which gives
the lhs of (\ref{Euler})\rbrack. As one can see, even an infinite number
of motion integrals does not fix the system but only its class.

Most authors feign indifference to the existence
of the infinite number of motion integrals in 2d turbulence.
Some arguments that only quadratic integrals
(i.e. energy and squared vorticity) should be taken into account
while considering thermodynamic equilibrium
were given by Kraichnan \cite{Kraa,Krad}.
However, an arbitrary turbulent pump generally produces a nonzero input
of all integrals $H_n$. The theory should describe the fate of these
integrals.

One could try to presume that $H_2$ cascades while other integrals
jump from the pumping to the damping and do not
influence the structure of the steady spectrum in the inertial interval
of scales. Let us show that this is impossible and that the
conditions for interaction locality are the same for any $n$.

The asymptotics of high-order correlation functions can be
expressed via $h_2(k)$ by the same means as asymptotics (\ref{as3}).
This can be done by using Wyld's diagram technique
which allows one to express any correlation function via series
containing the double correlation function and Green's function. For
$k\rightarrow0$, the double correlation function increases as
$k\sp{-x-y_2}$ while Green's function behaves as $k\sp{-x}$. Here $x$ is
a dynamic exponent, its value is irrelevant for our consideration.
Since we are dealing with the case $y_2>0$ the relevant terms in the
series are those which contain the double correlation function of small
wavenumbers \cite{LF92}. Such subsequences could be summed up to get
exact expression for the asymptotics of any correlation function.
Proceeding in this way we get the same convergence condition $y_2>2$ for
any $n$. In particular, Kraichnan's spectrum gives
the expression for $P_n$ containing $(\ln kL)\sp{n-1}$.
We should thus conclude that different vorticity integrals behave
in the same way in $k$-space.

The conservation of $H_n$ means that the Euler equation can be also
represented in the form
$\partial h_n(k)/\partial t +{\rm div}\,P_n=0$ so that the steady-state
condition takes the form $P_n=const$. Here the spectral density $h_n$
is expressed via the $n$-th correlator while the flux $P_n$ via the
$n+1$-th correlator. Condition of the flux constancy gives the exponent
of the $n$-th correlator $y_n=2n-2$ under the same condition of
interaction locality as (\ref{fc2}). Note that Kraichnan's spectrum
formally satisfies these conditions (formally but not really because of
the divergences). Indeed, the dimensions are additive in this case so
that estimating the rate of the viscous dissipation of $H_n$ ($n\geq2$)
similarly to (\ref{dissip}), one gets
\begin{equation}
\dot H_n=\nu\!\int\limits\sp{1/a}\!k\sp2h_n(k)\,dk\propto
\nu\sp{(n-1)(\Delta_\psi+1)+\Delta_\phi+2\over\Delta_\phi-\Delta_\psi}\
.\label{disint}\end{equation}
Kraichnan's spectrum gives $2\Delta\sb\psi=\Delta\sb\phi=-2$ so that
(\ref{disint}) is satisfied for any $n$. Indeed, this corresponds to
$\psi(r)\propto \vert r\vert\sp 2$ so that the vorticity $\omega$ has
zero scaling dimension (maybe logarithmic). All powers of the vorticity
can thus have constant fluxes in $k$-space simultaneously. Actually,
Kraichnan's spectrum would equally respect all conservation laws if it
was local.

Note that we estimated
the $n$-th correlation function in (\ref{disint})
in the simplest way just to show that $\Delta_\psi=-1$ formally satisfies
all conservation laws. As it is shown in the Appendix, a minimal model
cannot have $\Delta_\phi=2\Delta_\psi=-2$. Any conformal model that
satisfies Polyakov's conditions (\ref{fluxdim},\ref{regul})
gives some negative defect of dimensions while calculating correlation
functions. It means that the exponent in (\ref{disint}) is an upper
bound so that by virtue of (\ref{fluxdim}) one can get the inequality
$$\dot H_n<C\nu\sp{(n-2)(\Delta_\psi+1)/(\Delta_\phi-\Delta_\psi)}$$ with
some positive constant $C$. Therefore, for Polyakov's solution with $
\Delta_\psi<-1$, the integrals $H_n$ with $n>2$ are not dissipated in the
inviscid limit. Consider, for example, the flux of $H_3$: $\langle\dot
\omega\omega\sp2\rangle$. It is expressed via the correlator $\langle\phi
\psi\psi\rangle$. If we fuse $\phi$ and $\psi$ and take the most
relevant operator, say, $\chi$ so that the condition of the flux
constancy is $\Delta\sb\chi+3+\Delta\sb\psi+1=0$. For example, for
Polyakov's solution (2,21), $\Delta\sb\chi=-15/7$ so that the condition
of the flux constancy is not satisfied and
$dH_3/dt\rightarrow0$ as $\nu\rightarrow0$.

For the general case we need to calculate $\left<\dot\omega\omega\sp{n-1}
\right>$ which is given by $\left<\phi\psi\sp{n-1}\right>.$ If we take
$\psi\sp{n-1}$ to be the most relevant operator, say $\chi_{n-1}$, in the
OPE of $\underbrace{\lbrack\psi\rbrack\ldots\lbrack\psi\rbrack}_{n-1}$
then we get the condition for constant flux of $H_n$: $\Delta_\phi+2+
(n-1)+\Delta_{n-1}=0$, where $\Delta_n$ is the conformal dimension of
$\chi_n$. One can see that to satisfy these conditions for all $n$ one
needs infinite number of primary fields with negative dimension.

It was first pointed out in our preprint \cite{FH92} that a minimal
model cannot describe the spectrum with all vorticity fluxes. This was
then recognized by Polyakov \cite{pol2} who suggested that $H_n$ can
flow upscales for $n>2$ and downscales for $n=2$. This seems unlikely.
A general matching problem actually includes two integrals of motion:
energy $\int v\sp2dr$ and arbitrary functional of vorticity
$\int F(\omega)dr$. The conservation of the latter follows from Kelvin's
theorem on conservation of circulation. The integrals $H_n$ are the
Taylor coefficients of $F(\omega)$. It is natural to think that energy
conservation determines large-scale while Kelvin's theorem determines
small-scale turbulence. How one could divide Kelvin's theorem into two
parts? Higher vorticity integrals cannot cascade upscales since their
fluxes are incompatible with the inverse energy flux according to
Fjortoft's theorem. Indeed, to absorb a small value of a large-scale
vorticity one should simultaneously absorb a large value of the velocity
(that is of the energy). Finite energy flux into the
small scales means zero vorticity flux there.

Another possibility suggested in \cite{pol2} is to break the symmetry
assuming nonzero mean values of some fields. For fields with negative
dimensions this strongly modifies the correlation functions \cite{zamo}.
In particular, the condition $P_n=$const acquires the form $\Delta_\phi
+2+(n-1+\Delta_{n-1})-\Delta_{n+1}=0$.
This condition gives $\Delta_n\approx n\sp2/4$ at $n\gg1$ which is
incompatible with the assumption on negative dimensions.

Let us summarize. We have the set of conformal models which are at best
some particular solutions under an exotic condition of excitation.
Kraichnan's spectrum formally respects all conservation laws but it is
not a solution. How the true solution should look like?
There are two important points: the infrared
divergences and the presence of an infinite number of integrals
of motion. Unlike naive expectation, a true spectrum should drop with
$k$ faster than Kraichnan's one to compensate a nonlocal gain of
the integrals by the modes in the inertial interval directly from the
pump (cf. with \cite{FR92}). Power divergence that formally arises
can be considered in a spirit of a field theory approach \cite{FL92}
and be included in the flux value as well as ultraviolet divergences in
the theory of phase transitions are included in the true value of the
transition temperature. It means that we should find a small-scale
turbulent spectrum under the condition of the fixed value at pump scale.
This value just determines the flux. Nonlocality is thus inevitable in
any consistent approach to small-scale 2d turbulence.
Indeed, it is easy to show that local vorticity
cascade is impossible: if cascade was local, one can show that
the exponent of $n$-th correlator would be $y_n=2n-2$ which immediately
gives a logarithmic divergence. The positiveness of the anomalous
dimensions (which are the differences between true $y_n$ and 2n-2)
follows also from the inequality (see, e.g.
\cite{RS78}) $\mid v(l)-v(0)\mid\leq cl\ln l$ providing the absence
of singularities in 2d velocity field. Large vortices play an
important role in the direct vorticity cascade (see \cite{Lesi}
and references therein).

Kraichnan's spectrum together with higher correlation functions
having the exponents $y_n=2n-2$ seems to be a good bare solution
which should be subjected to the procedure of an ultraviolet
renormalization. Renormalization to be done should give the
anomalous dimensions linearly depending on the order of the
correlation function to provide for constant fluxes of all vorticity
integrals \cite{FL92}.

\leftline{\bf Acknowledgements: }
Discussions with P. Wiegmann, A. Finkelstein, G. Schutz, R. Plesser,
A. Schwimmer, Y. Levinson and V. Lebedev as well as useful remarks of
A. Polyakov are gratefully acknowledged. G. F. acknowledges support of
the Rashi Foundation.

\appendix
\section{}

Here we summarize all the conditions set by Polyakov
on the solutions of a turbulence problem.
We describe the algorithm we used in \cite{FH92} to compute
these solutions and give arguments why we think $\Delta_\psi=-1$ is
not a minimal model solution.

The conditions are:

1. $\Delta_\psi+\Delta_\phi=-3$

2. $\Delta_\psi<-1$

3. $\phi$ must be the operator with the smallest dimension in
the OPE of \lbrack$\psi$\rbrack\ and \lbrack$\psi$\rbrack.

The minimal models are characterized by two positive co-prime integers
$(p,q)$. For each minimal model $(p,q)$ there is a set of ${(p-1)(q-1)
\over 2}$ primary fields parameterized by two integers $(n,m)$, $1\leq
n\leq p-1$, $1\leq m\leq q-1$. The spectrum of conformal dimensions is
given by
\begin{equation}\Delta_{(n,m)}={(nq-mp)\sp2-(q-p)\sp2\over 4pq}\ .
\label{dnm}\end{equation}
{}From this formula we see that $\Delta_{(n,m)}=\Delta_{(p-n,q-m)}$ and
they both correspond to the same primary field. The scaling
dimension of the primary field $(n,m)$ is $2\Delta_{(n,m)}$.
{}From the first condition, using (\ref{dnm}),
we get ${p\over q}$ in terms of $n_\psi,m_\psi,n_\phi,m_\phi$
$${p\over q}={n_\psi m_\psi+n_\phi m_\phi-8+k\over m_\psi\sp2+
m_\phi\sp2-2},$$
where $k$ is an integer defined by
\begin{equation}k\sp2=(n_\psi m_\psi+n_\phi m_\phi-8)\sp2-(n_\psi\sp2+n_
\phi\sp2-2)(m_\psi\sp2+m_\phi\sp2-2). \label{ks}\end{equation}
This defines both $p$ and $q$ since they are co-prime. From the selection
rules for the OPE we have $n_\phi,m_\phi$ odd numbers and satisfy $1\leq
n_\phi\leq2n_\psi-1$, $1\leq m_\phi\leq2 m_\psi-1$. Requiring $\Delta_
\phi$ be minimal we get $(n_\phi q-m_\phi p)\approx0$, this sets
$m_\phi$ to be the nearest odd to
\begin{equation}n_\phi{n_\psi m_\psi-8\pm\sqrt{2m_\psi\sp2-16n_\psi
m_\psi+2m_\psi\sp2+60}\over n_\psi\sp2-2}. \label{mpha}\end{equation}
An algorithm to calculate these turbulent solutions can be as follows.
Given $n_\psi,m_\psi$, one can calculate $m_\phi$ from (\ref{mpha}) and
if $k$ from (\ref{ks}) is an integer one can also find $p$ and $q$. Using
this algorithm we got a set of solutions up to $n_\psi=10$ \cite{FH92}.

We also want to check if there exists a minimal model for which $\Delta_
\psi=-1$ or $\Delta_\phi=-2$. We set $\vert n_\phi q-m_\phi p\vert$ to
its minimal value and since $p,q$ are co-prime there exist $n_\phi,m_\phi
$ such that $n_\phi q-m_\phi p=1$ and we get, using (\ref{dnm}), $-2={1
-(p-q)\sp2\over4pq}$ or
\begin{equation}p\sp2-10pq+q\sp2-1=0.\label{pq}\end{equation}
for large $p,q$ one can neglect 1 and solve  the quadratic
equation to get
\begin{equation}\left({q\over p}\right)_\pm=5\pm\sqrt{24},\label{qop}
\end{equation}
the two solutions correspond to
the symmetry between $p$ and $q$. (This solution is irrational so in
practice one can take $p,q$ co-prime such that $p\over q$ is
close to this value, one can get any accuracy by taking large $p,q$).
{}From $\Delta_\psi=-1$ and (\ref{qop}) we have $m_\psi$ in terms of
$n_\psi$ as the nearest integer to
\begin{equation}
m_\psi=n_\psi{q\over p}+2\sqrt{q\over p}.\label{mn}\end{equation}
The sequence defined by
$$a_{k+1}=10a_k-a_{k-1}\qquad a_0=0,\quad a_1=1$$
serves as a solution to (\ref{pq}) taking $p=a_k$, $q=a_{k+1}$ for any
$k=2,3,...$, the $k$-th term is given by
$$a_k={(5+\sqrt{24})\sp k-(5-\sqrt{24})\sp k\over2\sqrt{24}}$$
Demanding that this $(p,q)$ model has $\Delta_\psi=-1$ leads to the
condition that $\sqrt{4pq+1}$ is an integer or, in terms of this sequence
, that $${(5+\sqrt{24})\sp{2k-1}+(5-\sqrt{24})\sp{2k-1}+14\over24}$$
is an integer squared, which is unlikely.

In our preprint \cite{FH92},
one can find the models with $(p,q)$ close to (\ref{qop})
and $(n_\psi,m_\psi)$ satisfying (\ref{mn}):
$(13,129)$, $(39,389)$, $(109,1082)$, $(232,2295)$ and $(69,686)$ with
$-\Delta_\psi={561\over559},{391\over389},{59001\over58969},{17873\over
17748},{7905\over7889}$ respectively.
\bigskip

\end{document}